\documentclass[aps,prl,reprint,twocolumn,showpacs,superscriptaddress,groupedaddress]{revtex4-1}  % for review and submission
\usepackage{graphicx}  % needed for figures
\usepackage{bm}        % for math
\usepackage{amssymb}   % for math
\usepackage{color}

\usepackage[hyperindex,breaklinks]{hyperref}

\begin{document}
%\linenumbers

\title{Search for Millicharged Particles Using Optically Levitated Microspheres}
\author{David C. Moore} \email{dcmoore@stanford.edu} 
\author{Alexander D. Rider} 
\author{Giorgio Gratta} 
\affiliation{Physics Department, Stanford University, Stanford, California 94305, USA} 

\date{\today}

\begin{abstract}
We report results from a search for stable particles with charge $\gtrsim
10^{-5}\ e$ in bulk matter using levitated dielectric microspheres in high
vacuum.  No evidence for such particles was found in a total sample of $1.4$~ng,
providing an upper limit on the abundance per nucleon of $2.5 \times 10^{-14}$
at the 95\% confidence level for the material tested.  These results provide the
first direct search for single particles with charge $\lesssim 0.1\ e$ bound in
macroscopic quantities of matter and demonstrate the ability to perform
sensitive force measurements using optically levitated microspheres in vacuum.

\end{abstract}

\maketitle

Millicharged particles, i.e., particles with charge $|q| = \epsilon e$ for
$\epsilon \ll 1$, have been proposed in extensions to the Standard Model that
include new, weakly coupled gauge sectors~(e.g.~\cite{Holdom:1986}).  It is
possible that millicharged particles are a component of the universe's dark
matter~\cite{Feldman:2007,McDermott:2011}.  If millicharged particles exist,
they could have been produced in the early universe~\cite{Chu:2012} and may have
formed stable bound states that can be searched for in terrestrial matter
today~\cite{Kim:2007,Langacker:2007}.

Constraints on millicharged particles exist from astrophysical and cosmological
observations~\cite{Davidson:2000, Dolgov:2013} as well as laboratory
measurements~\cite{Jones:1977,Prinz:1998,Bader:2007, Gninenko:2007, Perl:2009}.
However, such limits are typically sensitive to the mass of the particles and do
not provide significant constraints for particles with mass $m_\chi \gtrsim
1$~GeV~\cite{Davidson:2000}.  In contrast, searches for millicharged
particles bound in bulk matter are typically insensitive to the the mass of the
new particle and can probe masses $\gg 1$~GeV, but suffer from significant
uncertainty on the relic abundance in terrestrial materials. The terrestrial
abundance depends on the binding and ionization rate of the particles in the
early universe, as well as subsequent enrichment or depletion
processes~\cite{Langacker:2007}.

Previous bulk matter searches using magnetic
levitometers~\cite{LaRue:1981,Phillips:1988, Marinelli:1982, Smith:1989} or
high-throughput Millikan oil drop techniques~\cite{Kim:2007, Lee:2002,
  Joyce:1983, Polen:1987} focused on searching for free quarks with $\epsilon =
1/3$, and did not have sensitivity to single fractional charges with $\epsilon
\lesssim 0.1\ e$.  In this work we present results from a search for particles
with $\epsilon \gtrsim 10^{-5}\ e$ in bulk matter using optically levitated
microspheres in high vacuum~\cite{Yin:2013}.  At high vacuum, residual
dissipation of the microsphere motion from gas collisions becomes small, and
measuring the motion of the microsphere in three dimensions allows extremely
sensitive force detection~\cite{Geraci:2010,Yin:2013}.  Previous work has
demonstrated the trapping of microspheres in vacuum and cooling of the
center-of-mass motion, which is necessary to keep the microsphere stably trapped
at low pressures~\cite{Li:2011, Gieseler:2012, Keisel:2013}.

We have developed a system for trapping and cooling dielectric microspheres with
diameters $\gtrsim 1\ \mu$m in vacuum using a single, vertically-oriented laser
beam~\cite{Ashkin:1971, Li:2013}. This optical setup allows long working
distances between the focusing optics and trap location (few mm to few cm) and
active feedback cooling of the microsphere's motion through modulation of the
trapping laser intensity and position.  This cooling is used to stabilize the
microsphere in the trap once dissipation due to residual gas collisions is
insufficient to balance the heating of the microsphere's mechanical motion by
the laser~\cite{Li:2013}.

\begin{figure}[t]
\centering
\includegraphics[width=\columnwidth]{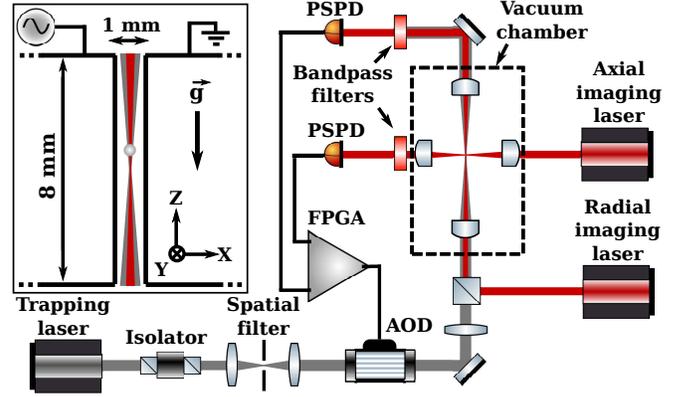}
\caption{Simplified schematic of the optical layout described in the text.  The
  inset (top left) shows a cross-section of the electrodes surrounding the
  trapping region. }
\label{fig:exp_setup}
\end{figure}

A schematic of the experimental setup is shown in Fig.~\ref{fig:exp_setup}.  A
300~mW diode laser ($\lambda = 1064$~nm) is spatially filtered and passes
through a 2-axis acousto-optic deflector (AOD).  Inside the vacuum chamber, the
beam is focused and recollimated by a pair of aspheric lenses with focal lengths
$f$=11~mm.  The resulting optical trap is centered between the flat faces of two
cylindrical electrodes with diameter $d = 8$~mm and separation $s =
1$~mm. Assuming an offset $< 100\ \mu$m from the center of the electrodes, this
configuration gives an expected electric field gradient at the microsphere
location $|\nabla \mathbf{E}/\mathbf{E}| \lesssim 4 \times
10^{-4}$~mm$^{-1}$~\cite{Atkinson:1983}.  In the following discussion, we define
X as the direction parallel to the electric field, while gravity points in the
-Z direction.

Two 20~mW diode lasers ($\lambda = 650$~nm) are used to provide 3D imaging of
the microsphere's position.  The first of these lasers is co-aligned with the
trapping beam after the AOD and passes through the same optics used to orient
and focus the trapping laser.  The transmitted light from this laser is imaged
onto a lateral effect position sensitive photodetector (PSPD)~\footnote{Thorlabs
  PDP90A, \protect\url{http://www.thorlabs.com}} to determine the
position of the microsphere in the 2 degrees of freedom (DOF) perpendicular to
the trapping beam axis (i.e. the X and Y axes).  The second imaging laser passes
through an orthogonal set of lenses and is used to image the position of the
microsphere along the Z axis.  A field programmable gate array (FPGA)-based
feedback loop modulates the amplitude and steering of the trapping beam using
the AOD, which allows feedback to be applied to all three orthogonal DOF.  An
additional lens is placed after the AOD such that deflections of the beam
angle at the AOD provide displacements of the trap~\cite{Neuman:2004}.

This work uses $5.06 \pm 0.44\ \mu$m diameter silica microspheres (with masses,
$m = 0.14 \pm 0.03$~ng) manufactured by Bangs
Laboratories~\footnote{\protect\url{https://www.bangslabs.com/}}.  Microspheres
are loaded into the trap in $\sim$50~mbar of dry nitrogen, where there is enough
gas damping that the trap is stable without feedback.  The microspheres are
applied to the bottom surface of a glass cover slip positioned above the
trapping region and a piezoelectric transducer is used to vibrate the cover
slip, releasing the microspheres to fall under
gravity~\cite{Ashkin:1971,Li:2013}.  A small fraction of these microspheres fall
through the trapping laser focus and become trapped.

Once a microsphere is trapped, the pressure in the chamber is reduced to 2~mbar
and feedback cooling is applied. The pressure in the chamber can then be lowered
to the base value of $2 \times 10^{-7}$~mbar.  We have demonstrated stable
trapping of a single microsphere at this pressure for more than 100~hr.
Although effective temperatures $< 10$~mK have been demonstrated for levitated
microspheres in vacuum in our setup and
elsewhere~\cite{Li:2011,Gieseler:2012,Keisel:2013}, cooling to $T_{eff} ~ \sim
1$~K is more than sufficient for the force measurements presented here. 

The minimum resolvable force in a 1~s integration is $\sigma_F = \sqrt{
  \sigma_{diss}^2 + \sigma_{im}^2 }$, where $\sigma_{diss} = \sqrt{4 k_B T_{eff}
  m \Gamma}$~\cite{Yin:2013,Li:2013} denotes the corresponding fluctuations due
to the total dissipation of the microsphere motion, $\Gamma$.  Here $k_B$ is
Boltzmann's constant and $m$ is the mass of the microsphere.  The additional
noise term, $\sigma_{im}$, denotes noise sources that are not associated with
damping, such as imaging noise.  When $\sigma_{im} \ll \sigma_{diss}$, the
application of feedback cooling does not significantly affect the
signal-to-noise of the measurement~\cite{Garbini:1996,Vitali:2002} since both
the signal and noise are attenuated by the feedback.  In this case, if the
dissipation, $\Gamma$, is limited by residual gas collisions, $\sigma_F \propto
\sqrt{P}$ for gas pressure $P$~\cite{Libb:2004}.  For our measurements the force
sensitivity is pressure limited for $P \gtrsim 10^{-3}$~mbar, while it is
limited to $\sigma_F = 5 \times 10^{-17}$~N~Hz$^{-1/2}$ at lower pressures by
laser fluctuations and imaging noise.  We are working to reduce these
non-intrinsic sources of noise and reach pressure limited noise at
$10^{-7}$~mbar, which is 2 orders of magnitude lower than the current
sensitivity.

The microspheres typically have a net charge of 100\textendash1000~$e$
after loading the trap. To remove this charge, a fiber-coupled Xenon
flash lamp is used to illuminate the electrode surfaces near the
microsphere. Empirically we have found that UV flashes from the Xe
lamp nearly always charge the microsphere in the negative direction,
consistent with photoelectric ejection of $e^-$ from the electrode
surfaces.  Hence, it is necessary to trap only microspheres with an
initial net positive charge.  When loading the microspheres directly
from the glass cover slip, we have found that the initial charge is
typically negative, but positive charges can be obtained by applying a
thin layer of plastic between the glass surface and the microspheres
(for these results, Scotch brand tape is used).

\begin{figure*}[htp]
\centering
\includegraphics[width=6.75in]{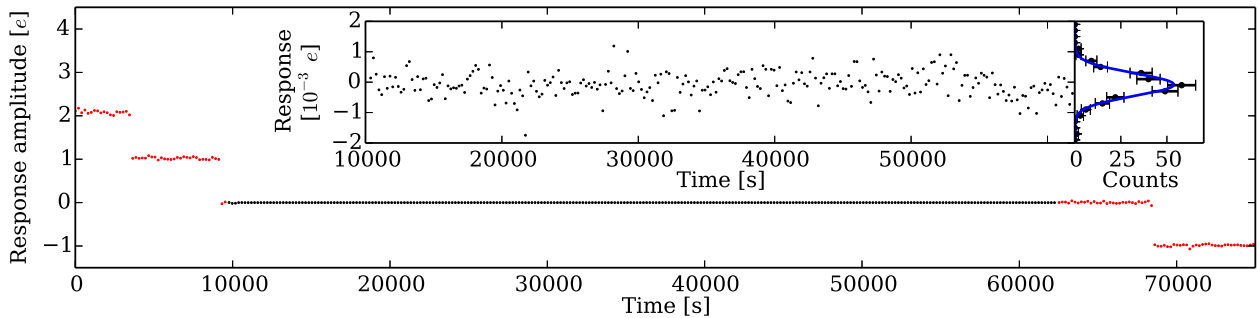}
\caption{(Color online) Measured position response in the X direction versus
  time as a microsphere is discharged.  The red (light) points show the
  calibration data at 10~V, while the black (dark) points indicate the data
  taken at 500~V with a net integral charge of 0~$e$. Each point represents the
  response to the drive measured in a single 100~s integration, after
  calibrating in units of $e$ using the discrete steps observed during
  discharging. (inset) Zoomed-in view of the nominally neutral data and a
  Gaussian fit to the residual response.}
\label{fig:discharging}
\end{figure*}

Once a microsphere is stably trapped at low pressure, an AC voltage with
$V_{peak} = 100$~mV and $f = 40$~Hz is applied to the electrodes. The charge of
the microsphere is inferred from its resulting motion while the Xe flash lamp is
used to quickly reduce the net charge to $\lesssim\ $+10~$e$.  While this coarse
discharging is done with a voltage drive at a single frequency, the data at $<
\pm 10\ e$ are taken with a broad frequency excitation to ensure that any
observed signal has the correct frequency response.  A frequency comb containing
all prime integers between 20 and 200~Hz is used, which has good frequency
coverage around the microsphere resonant frequency of $f \sim 150$~Hz.  Since
only prime frequencies are used, harmonics generated by potential
non-linearities can be identified.  After the microsphere has been discharged to
a net charge $\lesssim\ $+10~$e$, the drive voltage is increased to $V_{peak} =
10$~V, which is sufficient to observe individual steps of 1~$e$ in the
microsphere charge with high signal-to-noise, but low enough that the
microsphere is not pushed out of the trap (the measured trap depth is
$\sim$1~eV).  The microsphere is then slowly discharged to a net integral charge
of 0~$e$, while calibrating the signal amplitude at each charge step. Once the
microsphere is neutralized, the voltage is increased to $V_{peak} = 500$~V to
increase sensitivity to small residual charges, and data are
taken for a total integration time $\tau \approx 5 \times 10^4$~s.  Finally, the
voltage is again reduced to 10~V and additional calibration data are taken as
the microsphere is charged to $\sim -5\ e$.  An example of this discharging and
measurement process is shown in Fig.~\ref{fig:discharging}.

For each data set, the X, Y, Z, and sum signals from the PSPDs are digitized at
a sampling rate of 5~kHz using a 16-bit analog-to-digital converter (ADC).  The
drive signal is also directly recorded after passing through a resistive voltage
divider with a measured gain of $(4.6 \pm 0.1) \times 10^{-3}$, and digitized on
a separate 12-bit ADC to prevent electrical cross-talk to the position
signals. A common trigger is used to synchronize acquisition of the drive and
position signals, ensuring that the phase of the drive relative to the
start of the data acquisition is identical for each 100~s cycle of the frequency
comb.  Data from each cycle are individually recorded to disk.

Data were taken for 10~microspheres to quantify variations between microspheres
and to increase the total mass tested. The manual nature of the current loading
sequence makes it impractical to measure a substantially larger number of
microspheres.  For each microsphere, the calibration data taken at a residual
charge of 1~$e$ are used to determine the expected response to the drive signal,
which includes both the transfer function of the microsphere's motion in the
trap as well as any effects of the readout electronics on the measured position
response.  The amplitude of the response for each 100~s cycle is then estimated
using the optimal filter constructed from this response
template~\cite{Rabiner:1975}. This amplitude estimator is applied to the data at
each net charge, and the measured single $e$ steps in the response are used to
calibrate the residual response for each microsphere.  The amplitude estimator
is linear within 5\% for response amplitudes $\leq 5\ e \times 10$~V, so
non-linearity in the response is negligible for the smaller residual motion of a
neutral microsphere driven at 500~V.

\begin{figure}[htp]
\centering \includegraphics[width=3.6in]{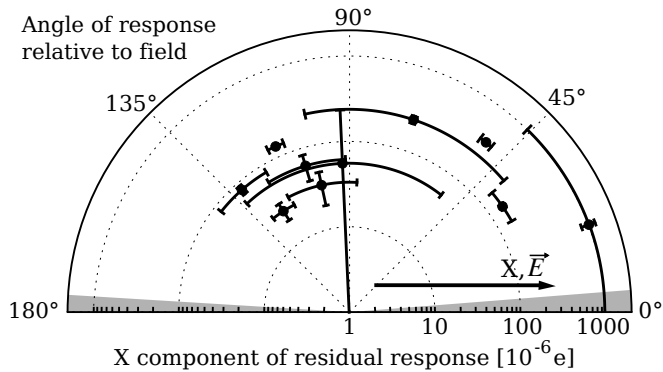}
\caption{Measured residual response in the direction of the electric field for
  each microsphere at a net integral charge of 0~$e$. Both the X component of
  the overall residual response as well as the smallest angle in 3D between the
  direction of the microsphere motion and the X~axis are shown. The gray band denotes
  the envelope of the measured response angles for all microspheres at net
  charges between $\pm 10\ e$.  While the microspheres show a statistically
  significant residual response, the measured angle of this response with
  respect to the field is typically inconsistent with the expected response for
  a residual net millicharge.}
\label{fig:bead_data}
\end{figure}

The residual response measured at a net integral charge of 0~$e$ for each of the
microspheres is shown in Fig~\ref{fig:bead_data}.  Roughly half of the
microspheres show a statistically significant residual of $(10 -
100)\times10^{-6}\ e$, while the remaining microspheres show a larger response
of $(100 - 1000)\times10^{-6}\ e$.  For the microspheres with measured residuals
$\lesssim 100 \times 10^{-6}\ e$, this response is found to vanish when the
electrodes are grounded at the input to the vacuum chamber, but all other wiring
connections are identical.  The residual also vanishes when the drive signal is
connected to the electrodes but there is no microsphere in the trap. These
checks indicate that the residual responses from $(10 - 100)\times10^{-6}\ e$
are likely to result from actual motion of the microspheres in the trap due to
the drive voltage.  However, this motion differs from the calibrated response of
a microsphere with a net charge $\sim 1\ e$, being typically offset by $45 -
90^\circ$ relative to the direction of the electric field.  In contrast, the
calibrations for all microspheres with charges between $\pm 10\ e$ align with
the nominal direction of the electric field within $< 5^\circ$, limited by
residual misalignment of the imaging axes to the field.  For the microspheres
with larger measured residuals, in some cases the residual response does not
completely vanish when the microsphere is removed from the trap, suggesting that
the residuals may be at least partly due to cross-talk from the drive signal to
the position measurement.

Given the misalignment of the microspheres' motion relative to the expected
response, the measured residuals are not evidence of a net residual fractional
charge.  Such residuals could be consistent with a permanent,
microsphere-dependent electric dipole moment that couples to the translational
DOF through asphericities or other inhomogeneities. Future work may allow the
reduction of this residual by spinning the microspheres around the Z-axis at
high frequency, averaging over inhomogeneities~\cite{Marinelli:1982}.  The
residual motion varies with each microsphere, and provides a systematic limit to
the sensitivity to small charges that dominates the statistical error for
integration times $\tau \gtrsim 10^3$~s.

\begin{figure}[htp]
\centering \includegraphics[width=\columnwidth]{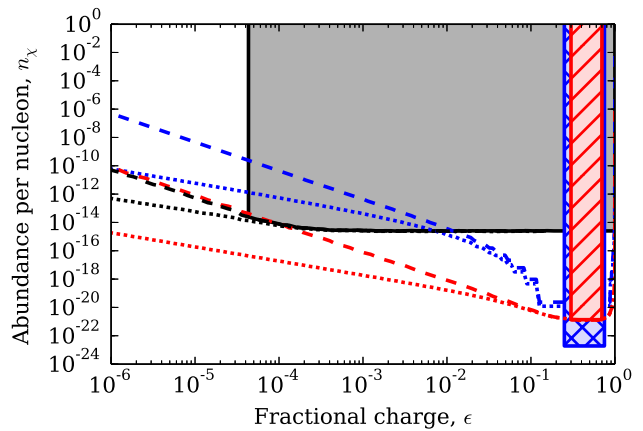}
\caption{(Color online) Limits on the abundance of millicharged particles per
  nucleon, $n_\chi$, versus the fractional charge, $\epsilon$.  The results from
  this work (black, solid fill) are compared to previous results from oil drop
  (blue, cross-hatched)~\cite{Kim:2007} and magnetic levitometer (red,
  hatched)~\cite{Marinelli:1982} experiments. The filled regions denote
  parameter space that is excluded at the 95\% CL for which a single particle of
  fractional charge $\epsilon$ could be observed by each experiment, and
  correspond to the published limits from Refs.~\cite{Kim:2007}
  and~\cite{Marinelli:1982}. The lines extending from each region show our
  calculation of the upper limits below the single particle threshold assuming
  either equal average numbers of positive and negative particles (dashed), or a
  single sign of particles (dotted).}
\label{fig:limit_plot}
\end{figure}

We conservatively calculate limits on the abundance of millicharged particles by
assuming that the component of the residual response in the direction of the
electric field could be due to a net fractional charge.  The resulting limit on
the abundance per nucleon, $n_\chi$, of particles with charge $ \pm \epsilon e$
is shown in Fig.~\ref{fig:limit_plot}, and compared to previous limits from oil
drop~\cite{Kim:2007, Lee:2002} and magnetic levitometer~\cite{LaRue:1981,
  Marinelli:1982, Smith:1989} experiments.  The gray shaded region denotes the
parameter space directly excluded by the present results assuming the residual
on each microsphere could be due to a single millicharged particle.

For the general case, where multiple millicharged particles could be bound to
the microspheres, the distribution of the number of millicharged particles per
microsphere is calculated for each value of $n_\chi$.  This distribution is
calculated assuming Poisson statistics and either that the average number of
positive and negative millicharged particles is equal (dashed lines in
Fig.~\ref{fig:limit_plot}), or that only particles of a single sign are trapped
in bulk matter (dotted lines in Fig.~\ref{fig:limit_plot}).  To be conservative,
the likelihood of a non-zero net charge is taken to be unity below the mean
residual measured for each microsphere, and to fall off with the measured
Gaussian error above the mean residual.  The likelihood marginalized over the
distribution of the number of millicharged particles is then calculated, and a
combined likelihood, $L_{tot}$, over all microspheres is formed.  The 95\%
confidence level (CL) upper limit for $n_\chi$ is determined from $L_{tot}$
following Wilks' theorem~\cite{Wilks:1938}.

Although Refs.~\cite{Kim:2007} and~\cite{Marinelli:1982} published limits only
above their single particle thresholds of $\epsilon \gtrsim 0.1$, we have
calculated the analogous limits from these experiments below this threshold for
comparison.  For Ref.~\cite{Marinelli:1982}, we use the reported residual charge
and errors for each levitated ball to calculate the limit following the same
technique as for our data above.  For Ref.~\cite{Kim:2007}, the total mass was
divided into $\sim 4.3 \times 10^7$ individual droplets, and only the
distribution of the measured residuals for these droplets was reported.
Following the procedure in~\cite{Kim:2007}, we calculate the 95\%~CL limit on
$n_\chi$ from this distribution given that no droplets were observed with $|q| >
0.25\ e$.  As shown in Fig.~\ref{fig:limit_plot}, the limits below the single
particle threshold fall off more quickly for Ref.~\cite{Kim:2007} than for
Ref.~\cite{Marinelli:1982} or our results.  This occurs because the total mass
tested in~\cite{Kim:2007} was divided into many small droplets, which reduces
the probability that a single droplet will contain multiple millicharged
particles.

As shown in Fig.~\ref{fig:limit_plot}, the present results provide the first
direct search for millicharged particles in bulk matter with sensitivity to
single particles with $5 \times 10^{-5} < \epsilon < 0.1$. Over this full range,
the upper limit on the abundance per nucleon is at most $n_\chi < 2.5 \times
10^{-14}$ at the 95\% CL for the material tested, while it improves to $n_\chi <
3.6 \times 10^{-15}$ for $\epsilon \gtrsim 10^{-3}$.  For $\epsilon < 5 \times
10^{-5}$, constraints on the abundance per nucleon are less stringent since
multiple millicharged particles per microsphere would be required to give an
observable charge.  Previous results from~\cite{Marinelli:1982} can also
constrain some or all of this region, depending on the assumed ratio of positive
to negative particles.  However, given the residual systematic effects seen in
all such searches, the single particle sensitivity of this work is necessary to
distinguish a signal from backgrounds since the expected discrete distribution
of particles can be directly observed.  Future work to reduce the systematics
described above may significantly improve the sensitivity of this technique.

In addition to providing the first direct search for single millicharged
particles with $\epsilon < 0.1$ in bulk matter, this work also represents the
first application of sub-aN force sensing using levitated microspheres in vacuum
to search for new particles or interactions. The method demonstrated here for
discharging microspheres with single-electron precision is applicable to future
work using optically levitated microspheres since it provides an absolute
calibration of the force sensitivity and a means of eliminating electrostatic
backgrounds that depend on net charge. In particular, these techniques and
further improvements may also allow significant increases in sensitivity to
short range forces~\cite{Yin:2013,Geraci:2010}, including searches for
non-Newtonian gravitational forces at micron distances.

This work was performed with seed funding from Stanford University.  We would
like to thank M.~Lu, M.~Dolinski, and M.~Celentano for their work on early
versions of the apparatus discussed here.  We would also like to thank
P.~Graham, J.~Mardon, and S.~Block for useful discussions.

\bibliography{millicharge_PRL}

\end{document}